\documentclass[twocolumn,amsmath,amssymb,aps, prl,
showkeys,showpacs, superscriptaddress, floatfix] {revtex4-2}
\usepackage{graphicx}
\usepackage[usenames,dvipsnames]{color}
\usepackage{bm}
\usepackage[colorlinks=true,breaklinks=true,allcolors=blue]{hyperref}
\usepackage{dcolumn}
\usepackage{slashed}
\usepackage{mathtools}
\usepackage{multirow}

\allowdisplaybreaks

\begin{document}
\title{Entanglement entropy in the Ising model with topological defects }
\author{Ananda Roy}
\email{ananda.roy@physics.rutgers.edu}
\affiliation{Department of Physics and Astronomy, Rutgers University, Piscataway, NJ 08854-8019 USA}
\author{Hubert Saleur}
\affiliation{Institut de Physique Th\'eorique, Paris Saclay University, CEA, CNRS, F-91191 Gif-sur-Yvette}
\affiliation{Department of Physics and Astronomy, University of Southern California, Los Angeles, CA 90089-0484, USA}

\begin{abstract}
Entanglement entropy~(EE) contains signatures of many universal properties of conformal field theories~(CFTs), especially in the presence of boundaries or defects. In particular, {\it  topological} defects are interesting since they reflect internal symmetries of the CFT, and have been extensively analyzed with field-theoretic techniques with striking predictions. So far, however, no lattice computation of EE has been available. Here, we present an ab-initio analysis of EE for the Ising model in the presence of a topological defect. While the behavior of the EE depends, as expected, on  the geometric arrangement of the subsystem with respect to the defect, we find that zero-energy modes give rise to crucial finite-size corrections. Importantly, contrary to the field-theory predictions, the universal subleading term in the EE when the defect lies at the edge of the subsystem arises entirely due to these zero-energy modes and is not directly related to the modular S-matrix of the Ising CFT. 
\end{abstract}
\maketitle 

Entanglement plays a central role in the development of long-range correlations in quantum critical phenomena. Thus, quantification of the entanglement in a quantum-critical system provides a way to characterize the universal properties of the critical point. The von-Neumann entropy is a natural candidate to perform this task. For zero-temperature ground-states of 1+1D quantum-critical systems described by conformal field theories~(CFTs), the von-Neumann entropy~[equivalently, entanglement entropy~(EE)] for a subsystem exhibits universal logarithmic scaling with the subsystem size~\cite{Holzhey1994, Calabrese2004}. The coefficient of this scaling determines a fundamental  property of the bulk CFT:  the central charge. For finite systems with boundaries at a conformal critical point, the EE receives universal, subleading, boundary-dependent contributions, the so-called `boundary entropy'~\cite{Affleck1991} -- a central concept in a variety of physical problems both in condensed matter physics and string theory. This boundary contribution to the EE provides a valuable diagnostic for identifying the different boundary fixed points of a given  CFT~\cite{Calabrese2004, Calabrese2009, Affleck2009, Roy2020a,Roy2020b}. 

While entanglement has been analyzed extensively in CFTs with and without boundaries, its behavior is much less understood in the presence of defects. The question is particularly intriguing since entanglement measures may provide an alternate way to classify defects in CFTs. Of particular interest are topological~(perfectly-transmissive) defects~\cite{Petkova2000, Bachas2001, Frohlich2004, Frohlich2006,Aasen2016}.  These defects commute with the generators of conformal transformations and thus, can be deformed without affecting the values of the correlation functions as long as they are not taken across field insertions~(hence the moniker topological). They reflect the internal symmetries of the CFT and relate the order-disorder dualities of the CFT to the high-low temperature dualities of the corresponding off-critical model~\cite{Frohlich2004, Krammers1941,Savit1980}. They also play an important role in the study of anyonic chains and in the correspondence between CFTs and three-dimensional topological field theories~\cite{Buican2017}.

It is natural to analyze EE in the presence of topological defects. Two distinct geometries have been considered in the literature: i) where the defect is entirely within the subsystem, and  ii) where the defect is located precisely at the interface between the subsystem and the rest. While both cases exhibit identical leading-order logarithmic scaling with subsystem size, they differ when subleading, {\it i.e.}, $O(1)$, corrections are taken into account. In the first case, after usual folding maneuvers, the subleading term can be equated to a boundary entropy with double the bulk degrees of freedom~\cite{Oshikawa1997, Saleur1998, Saleur2000}. In this way, the subleading correction to symmetric EE can be computed  analytically for all rational CFTs~\cite{Gutperle2015}. In the second case, the subleading term for the interface EE is much more difficult to obtain. Field-theory computations based on the replica trick for the free, real boson~\cite{Sakai2008}, the free, real fermion~\cite{Brehm2015}~(see also~\cite{Eisler2010, Peschel2012e, Calabrese2011ru}) and generalization to all rational CFTs~\cite{Gutperle2015} using the corresponding twisted torus partition functions~\cite{Petkova2000} provide results whose validity have never been tested with ab-initio, lattice computations. Such a computation is particularly important since the mapping to the twisted torus partition function does not faithfully capture the geometric arrangement of the subsystem with respect to the defect. This is important for subleading terms in the interface EE which are, in fact, the signatures of the topological defect. 

Our goal, in this paper, is to investigate the question in the simplest possible case, the Ising model, using ab-initio calculations. These calculations are more complex than is usually the case for EE due to the presence of zero-energy modes. The latter are a salient feature of the topological nature of the defect. While the effects of zero-modes have been extensively quantified in gapped systems due to their relevance in topologically ordered systems, the same is much less understood in the critical systems. Here, we show that these zero-modes give rise to nontrivial contributions to the EE when the subsystem size is comparable to the size of the whole system, similar to the case of a periodic ring of free fermions~\cite{Herzog2013, Klich2017}. Thus, these zero-modes profoundly affect corrections to scaling. After the role of these zero-modes is properly taken into account, we find  results that are as expected for case (i), but are not compatible with the formulas proposed in the string/field theory literature~\cite{Brehm2015,Gutperle2015} for case (ii). 
\begin{figure}
\centering
\includegraphics[width = 0.49\textwidth]{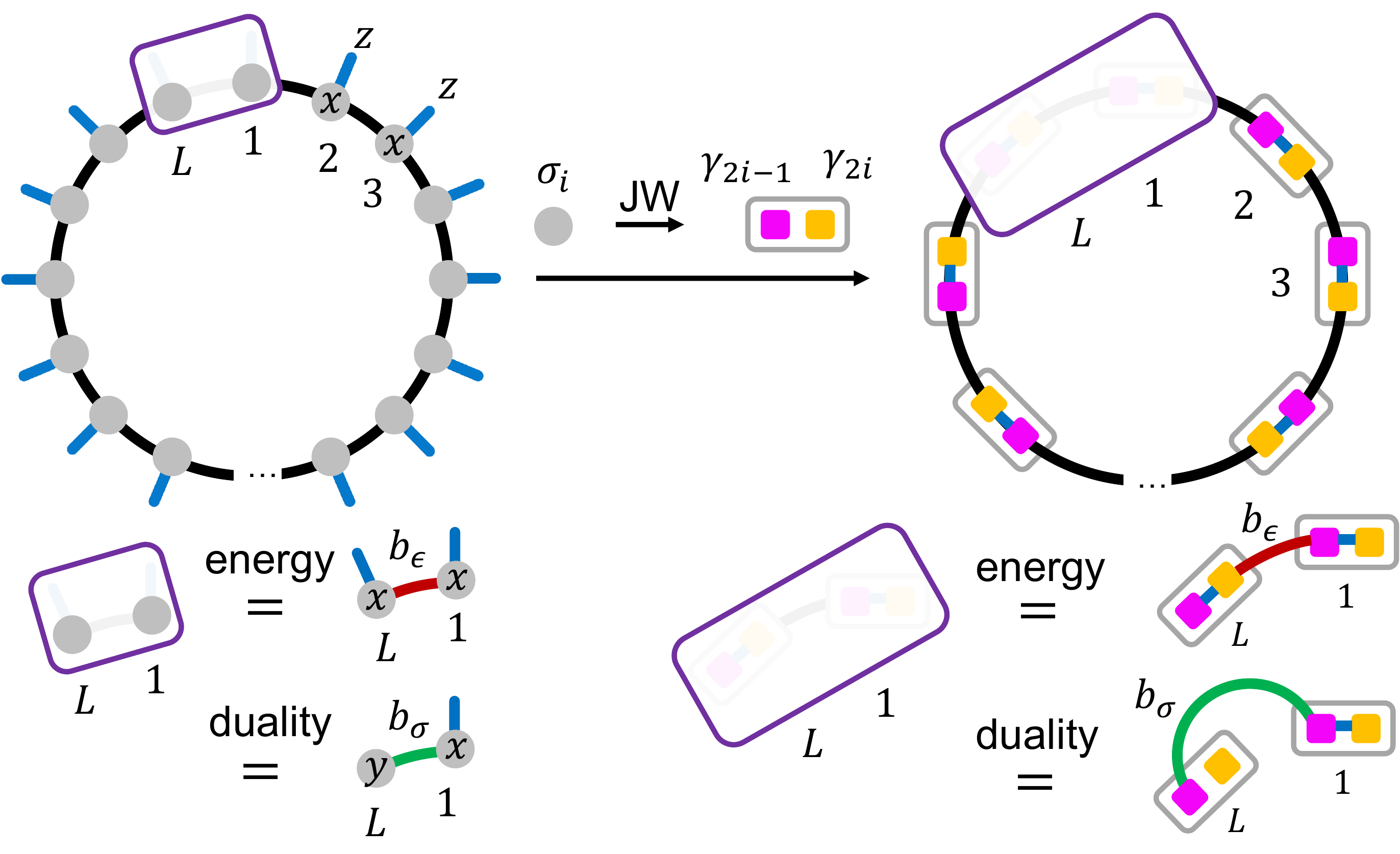}
\caption{\label{defect_schematic} Schematic of a single defect in the critical Ising chain with periodic boundary conditions in the spin~(left) and fermionic~(right) picture. Away from the defect, the Hamiltonian contains nearest-neighbor ferromagnetic~$\sigma_i^x\sigma_{i+1}^x$ interaction~(black lines) and onsite transverse field~$\sigma_i^z$~(blue lines). The defect~(purple box) concerns the spins located at sites $L$ and 1. The energy~(duality) defect part of the Hamiltonian is: $b_\epsilon\sigma_L^x\sigma_1^x/2$ $+ \sigma_L^z/2$~($b_\sigma\sigma_L^y\sigma_1^x/2$). After Jordan-Wigner~(JW) transformation, the ferromagnetic coupling corresponds to an interaction of the form~$\gamma_{2i}\gamma_{2i+1}$, while the transverse field corresponds to~$\gamma_{2i-1}\gamma_{2i}$. For the duality defect, there is a zero-mode localized at $\gamma_{2L}$. Another zero-mode is delocalized throughout the system~(see below). }
\end{figure}

The Ising model provides a particularly appropriate testbed for the EE in the presence of topological defects due to the following. First, the lattice Hamiltonians are well-understood~\cite{Grimm2001,Oshikawa1997}. Second, the zero-mode structure is the simplest and yet, sufficient to explain the main concepts. Third, the defect Hamiltonian can be mapped, after Jordan-Wigner~(JW) transformation, to a bilinear fermionic Hamiltonian. The latter can be diagonalized semi-analytically and leads to very accurate predictions of EE~\cite{Vidal2002,Peschel2003, Latorre2004}. To emphasize the nontrivial nature of the topological defect, we compare our results with the non-topological defects in the model. 

There are two nontrivial classes of defects in the Ising CFT: energy~($\epsilon$) and duality~($\sigma$). We consider a periodic Ising chain with the defect concerning the spins at sites $L$ and 1~\cite{Henkel1989, Baake1989, Grimm2001}. The Hamiltonians are:  $H_{\epsilon, \sigma} = H_0$$- h_{\epsilon, \sigma}$~(see Fig.~\ref{defect_schematic}). Here,  the bulk Hamiltonian term, $H_0, = -\sum_{i=1}^{L-1}\sigma_i^x\sigma_{i+1}^x/2$ $- \sum_{i = 1}^{L-1}\sigma_i^z/2$. The two defect terms are: $h_\epsilon = b_\epsilon\sigma_L^x\sigma_1^x/2$ $+ \sigma_L^z/2$ and $h_\sigma = b_\sigma\sigma_L^y\sigma_1^x/2$~\footnote{Note that $h_{\epsilon,\sigma}$ are marginal perturbations which, in general, lead to continuously varying critical exponents~\cite{Cardy1987}.}. The energy defect consists of one bond with altered strength $b_\epsilon$ connecting spins at sites $L$ and 1. In particular, $b_\epsilon = 0, +1$ and $-1$ correspond to Ising models with open, periodic and antiperiodic boundary conditions~(bcs) respectively. Unlike the energy defect, the duality defect consists of a $\sigma_L^y\sigma_1^x$ interaction~\footnote{This duality defect Hamiltonian is related by a local unitary rotation on the $L^{\rm th}$ spin to the one considered in Refs.~\cite{Oshikawa1997}, which has $\sigma_L^z\sigma_1^x$ interaction. We do not use this alternate form since it no longer leads to a bilinear Hamiltonian under JW transformation and cannot be solved by free-fermion techniques. }. Equally important, there is no transverse field at the $L^{\rm th}$ site. The duality defect for $b_\sigma = 1$~(equivalently $b_\sigma = -1$, which is related by a local unitary rotation) is the topological defect for the Ising CFT. 

Next, we perform a JW transformation: $\gamma_{2k-1} = \sigma_k^x\prod_{j=1}^{k-1}\sigma_j^z$, $\gamma_{2k} = \sigma_k^y\prod_{j=1}^{k-1}\sigma_j^z$, where $\gamma_{j}$-s are real, Majorana fermion operators obeying $\{\gamma_j, \gamma_k\} = 2\delta_{j,k}$. In the fermionic language, the defect Hamiltonians are  $H_{\epsilon, \sigma}^f = H_0^f - h_{\epsilon, \sigma}^f$, where
\begin{align}
\label{H_0_f}
H_0^f &=  \frac{i}{2}\sum_{j=1}^{L-1}\gamma_{2j}\gamma_{2j+1} + \frac{i}{2}\sum_{j=1}^{L-1}\gamma_{2j-1}\gamma_{2j},\\\label{dH_f}
h_\epsilon^f &=  \frac{ib_\epsilon}{2}\gamma_{2L}\gamma_1 - \frac{i}{2}\gamma_{2L-1}\gamma_{2L},\ h_\sigma^f = -\frac{ib_\sigma}{2}\gamma_{2L-1}\gamma_1.
\end{align}
Here we have restricted ourselves to the symmetry sector $Q=\prod_{j=1}^L\sigma_j^z = 1$~\footnote{Equivalently, we analyze the mixed sector Hamiltonian~\cite{Grimm2001}: $H_{\epsilon,\sigma}^m = P_+H_{\epsilon, \sigma}^f(Q = +1) +  P_-H_{\epsilon, \sigma}^f(Q = -1)$}. For the energy defect, from the definition of $H_\epsilon^f$, we recover the well-known fact that the periodic~(antiperiodic) coupling at the boundary for the spin model corresponds to antiperiodic~(periodic) coupling for the fermionic model. In particular, the periodic fermionic model~($b_\epsilon = -1$) contains two nonlocal Majorana zero-modes, which together are responsible for the two-fold degenerate ground-state of the fermionic model. For the duality defect, the operator $\gamma_{2L}$ does not occur in $H_\sigma^f$. It commutes with the Hamiltonian: $[\gamma_{2L}, H_\sigma^f] = 0$, and anticommutes with the conserved $\mathbb{Z}_2$ charge: $\{\gamma_{2L},Q\} =0$. Thus, it is a zero-mode of the model which is perfectly localized in space. It has a partner zero-mode which is completely delocalized: $\Lambda(b_\sigma) = \sum_{k=1}^L\gamma_{2k-1} + b_\sigma\sum_{k = 1}^{L-1}\gamma_{2k}$. Note that the zero-modes exist for all values of $b_\sigma$ and are not special features of the topological point. The fermionic Hamiltonian also reaffirms a CFT result~\cite{Grimm2001}:~$H_\sigma^f$ describes a chain of $2L-1$ Majorana fermions or equivalently, $L-1/2$ spins. This is important for quantifying finite-size effects. 

Now, we compute symmetric and interface EEs for the ground states of the fermionic Hamiltonians $H_{\epsilon, \sigma}^f$. Since the latter are bilinear in the fermionic operators, we compute the relevant EEs from the ground-state correlation matrix~\cite{Vidal2002,Peschel2003, Latorre2004}. The latter is calculated from the ground state by filling up the negative energy states. The method is unambiguous in the absence of zero-energy modes. However, in the presence of the latter~({\it e.g.}, $b_\epsilon  = -1$ or any $b_\sigma$), it raises the question: are the zero-energy states empty or occupied in the ground state? Yet another possibility is to consider an incoherent superposition of filled and empty states. This leads to the total system being in a mixed state, but is appropriate when taking the zero-temperature limit of a thermal ensemble~\cite{Herzog2013}. The question is crucial to the computation since zero-energy modes nontrivially affect the EE. For $b_\epsilon = -1$, the zero-modes give rise to nontrivial corrections to the EE of a subsystem of size~$r$ within a total system of size~$L$~\cite{Herzog2013, Klich2017}. The correction~$\Delta S(r/L) =$ $\frac{\pi r}{L} \int_0^\infty \tanh(\pi r h/L)[\coth(\pi h) - 1]$. For $r\ll L$, the EE is oblivous to the existence of the two nonlocal zero-modes spread throughout the system:~$\Delta S\simeq \pi^2r^2/12L^2\rightarrow 0$. The situation changes as the subsystem occupies appreciable fraction of the total system~($r\sim L$) culminating in $\Delta S(r = L) = \ln2$, the latter being the entropy of the two-fold degenerate ground state of a periodic chain of fermions. Below, we present analogous results for the topological defect  considering separately the cases when the total system is in a pure and mixed state for the relevant fermionic models~\footnote{See Supplementary Material for details}. 
\begin{figure}
\centering
\includegraphics[width = 0.49\textwidth]{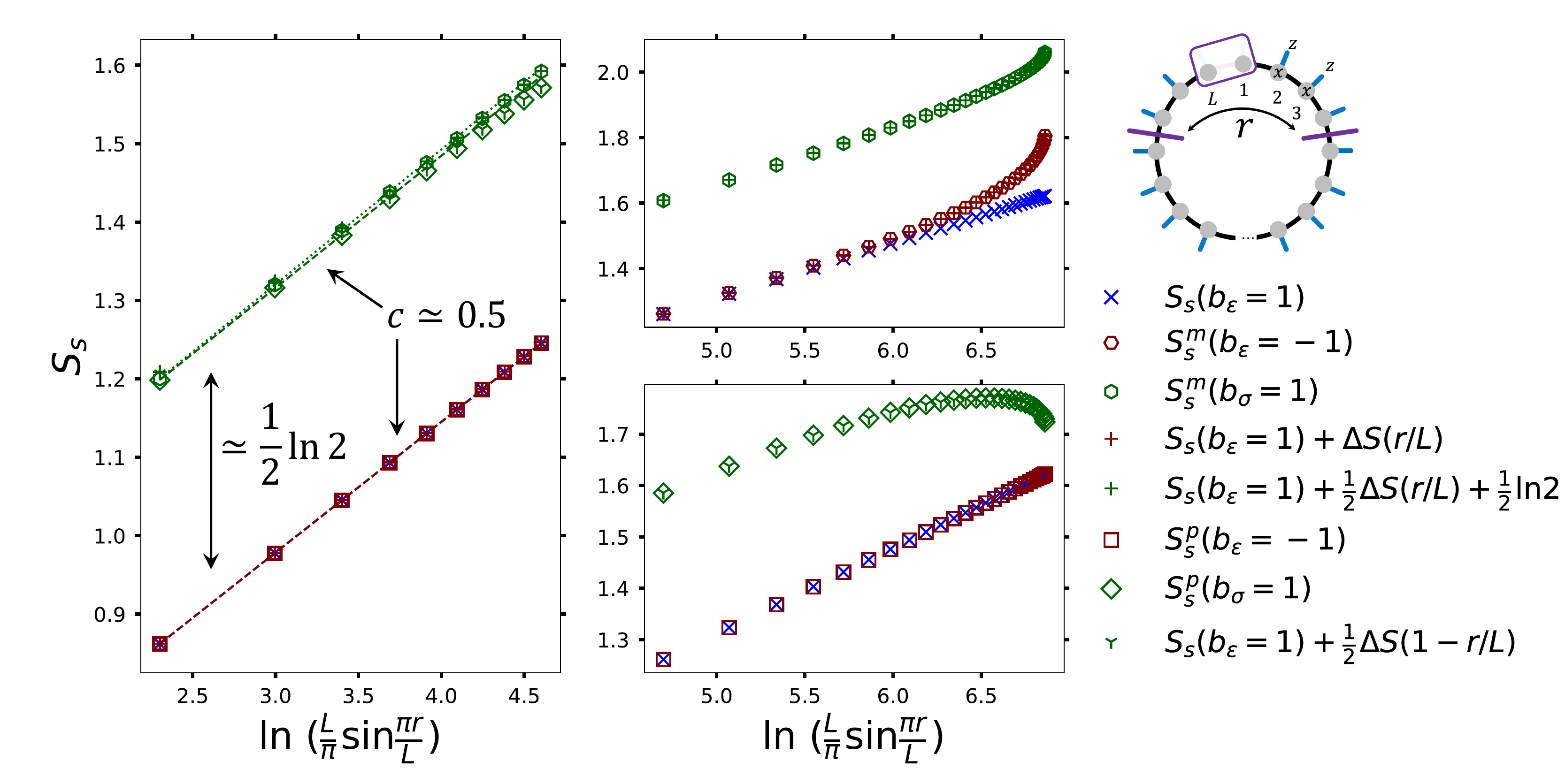}
\caption{\label{symm_entropy} Results for the symmetric EE~($S_s$) for a periodic chain of size~$L = 3000$ with a single defect and $10<r<100$~(left panel) and $100<r<L/2$~(right panels). The blue crosses are obtained for~$b_\epsilon = 1$~({\it i.e.}, no defect). The maroon~(green) hexagons are for the energy defect~$b_\epsilon = -1$~(duality defect~$b_\sigma = 1$), when the total system is in a mixed state. The differences with $S_s(b_\epsilon=1)$ are $\Delta S(r/L)$ and $\Delta S(r/L)/2 + (\ln2)/2$ for the two cases. The corresponding predictions are denoted by maroon and green pluses. Compared to~$b_\epsilon = -1$, the EE for~$b_\sigma = 1$ has an additional offset $(\ln2)/2$ even for $r/L\rightarrow0$ due to the zero-mode $\gamma_{2L}$ localized at the center of the interval. As $r/L$ increases, the EE for~$b_\sigma = 1$, due to the single nonlocal zero mode,~$\Lambda(b_\sigma=1)$, receives a contribution half as large as that for $b_\epsilon = -1$, which has two such modes. The maroon squares~(green diamonds) present the results when the total system is in a pure state~(the zero-energy state being filled or empty) as opposed to a mixed state. The results for $b_\epsilon = \pm1$ are indistinguishable. However, relative to the energy defect, the $b_\sigma=1$ shows an offset~$\Delta S(1-r/L)/2$, which again reduces to~$(\ln2)/2$ for $r/L\ll1$. }
\end{figure}
\begin{figure}
\centering
\includegraphics[width = 0.49\textwidth]{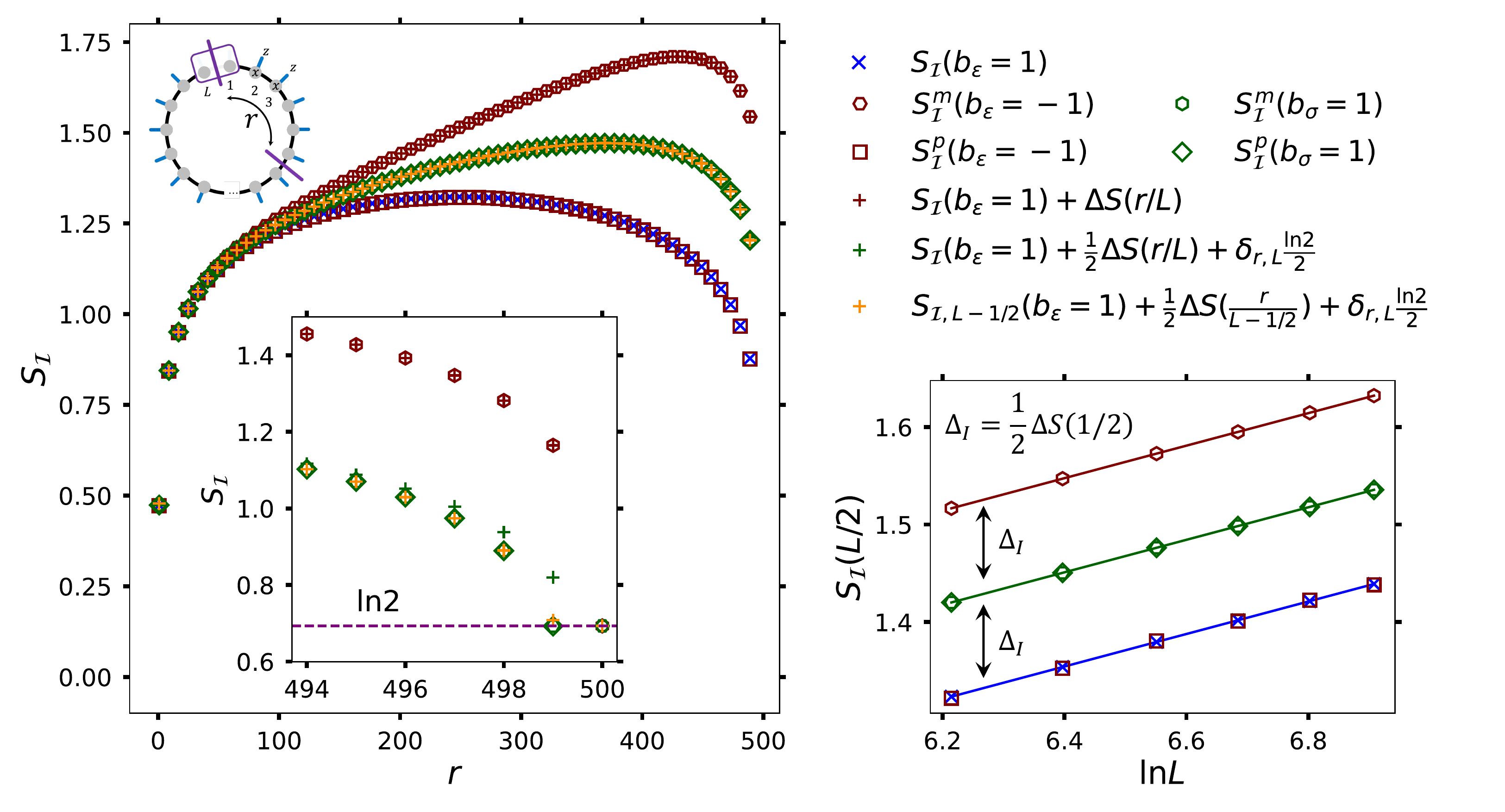}
\caption{\label{inter_entropy} (Left panel) Results for the interface EE~($S_{\cal I}$) for a periodic chain of $L = 500$ with a single defect. One end of the subsystem is at the defect and the other end sweeps through the system. The~$b_\epsilon=-1$ case, both when the system is in an incoherent superposition~(maroon hexagons) and a pure state~(maroon squares), can be directly understood from the symmetric EE~(Fig.~\ref{symm_entropy}). The topological defect results are identical for mixed~(green hexagons) and pure~(green diamonds) total-system state. Then, the interface EE is best compared to that of a spin-chain of size $L-1/2$ without defects~[see Fig.~\ref{defect_schematic} and discussion below Eq.~\eqref{H_0_f}]. The difference between the two EEs is $\Delta S[r/(L-1/2)]+\delta_{r,L}(\ln2)/2$~(orange pluses). For comparison, the corresponding predictions obtained by comparing to size-$L$ chain without defect~(green pluses) are also plotted. For $r$ not too close to $L$, the difference is negligible. But, zooming in around $r\sim L$~(inset) makes the difference manifest.~(Right panel) We show the scaling of the interface EE for $r=L/2$ as a function of $\ln L$. Every curve exhibits the same leading-order scaling yielding a central charge $c\simeq 0.5$. However, the offset with respect to no-defect for ~$b_\epsilon=-1$ is $\Delta S(1/2)=-1/2 + \ln2$ when the total system is mixed and 0 when pure. On the other hand, the corresponding offsets for both pure and mixed cases for $b_\sigma=1$ are $\Delta S(1/2)/2=-1/4 + (\ln2)/2$.}
\end{figure}

First, we compute the symmetric EE~(Fig.~\ref{symm_entropy}). The results for~$b_\epsilon = 1$~(no defect) are shown with blue crosses. The symmetric EE exhibits the expected logarithmic scaling: $S_s(b_\epsilon = 1)$ $= \frac{c}{3}\ln\big[\frac{L}{\pi}\sin\big(\frac{\pi r}{L}\big)\big]$ $+ S_0$ for all values of $r/L$. We set the lattice spacing to 1 throughout this work. Fitting this expression yields the expected central charge~$c\simeq0.5$ and~$S_0\simeq0.478$. The maroon~(green) hexagons corresponds to the symmetric EE for $b_\epsilon = -1$~($b_\sigma =1$) when the total system is in an incoherent superposition of the zero-energy states being filled and empty. Compared to $S_s(b_\epsilon=1)$, the symmetric EEs~($S_s^m, m$ denoting the total system being mixed) for the two cases get an additional contribution of $\Delta S(r/L)$~\cite{Klich2017} and $\Delta S(r/L)/2+ (\ln2)/2$. 
Thus, when $r/L\ll1$, for both $b_\epsilon=-1$ and $b_\sigma=1$, $S_s^m$ exhibits the expected logarithmic dependence with $c\simeq0.5$~(left panel). However, the offset,~$S_0$, for $S_s^m(b_\sigma=1)$ is $(\ln2)/2$ higher. This higher offset unambiguously distinguishes the topological defect from the energy defect and is because of the {\it localized} unpaired Majorana zero mode at the center of the subsystem~(the result will be the same as long as the defect lies within and not at the edge of the subsystem). This is consistent with the identification of this defect problem to a boundary CFT problem at the `continuous Neumann boundary fixed-point' after folding, with the corresponding $g$-function $=\sqrt{2}$~\cite{Oshikawa1997}~(see also Ref.~\cite{Alba2017}). This should be compared with the `continuous Dirichlet boundary fixed point'~($b_\epsilon=-1$ case), which has $g$-function $=1$ and thus, no additional boundary entropy contribution. For~$b_\sigma=1$, increasing $r/L$  leads to a further offset of $\Delta S(r/L)/2$ due to the contribution from the second {\it nonlocal} zero-mode. Here, the factor of 1/2 accounts for the difference in the number of nonlocal zero-modes in the $b_\epsilon=-1$ and $b_\sigma=1$ models. The maroon squares~(green diamonds) correspond to the symmetric EEs obtained by keeping the zero-energy state empty~(denoted by~$S_s^p, p$ denoting the total system being pure; the results are identical for the filled case).  Now, the results for $b_\epsilon= \pm1$ are indistinguishable. However, compared to the case without defects, $S_s^p(b_\sigma = 1)$ exhibits a $\Delta S(1-r/L)/2$ offset. For $r\ll L$, this again leads to the offset, $S_0$, being $(\ln2)/2$ higher than the other cases. For $r\sim L$, the $S_s^p$ diminishes compared to the case when the total system is mixed due to the purity of the total system. 

Next, we compute the interface EE~(Fig.~\ref{inter_entropy}), where one end of the subsystem~(of size $r$) is located at the defect and the other end sweeps through the system~(of size $L$)~\footnote{Note that the symmetric EE computation required much larger total system sizes in order to get a good result for offset $(\ln2)/2$.}. For $b_\epsilon = -1$, the results are identical to the symmetric case. This is expected since the resulting model is just a fermionic model with periodic bc and there is no difference between symmetric and interface EEs. The results for $b_\sigma=1$ are the same for the total system in a mixed~(green hexagons) and pure~(green diamonds) state. Then, $S_{\cal I}^{m,p}(b_\sigma=1)$ is given by the EE of an Ising chain of length $L-1/2$ without any defects~[$S_{{\cal I}, L-1/2}(b_\epsilon=1)$] together with an offset. The first contribution to this offset is $\Delta S[r/(L-1/2)]$. It arises as the subsystem size grows and becomes aware of the nonlocal zero-mode~$\Lambda(b_\sigma)$ in the chain of length~$L-1/2$. The second contribution, $\delta_{r,L}(\ln2)/2$, arises only when $r=L$. This is due the localized zero-mode, $\gamma_{2L}$, which contributes only when the subsystem covers the entire system. The resulting prediction is shown with orange pluses. For comparison, we have also shown the curves obtained by computing the corresponding offsets for the system-size $L$~(green pluses). For $r$ not close to $L$, both predictions work well. However, for $L-r\sim 1$, only the computation with system-size $L-1/2$ leads to the correct predictions~(see inset). The field-theory computations are usually done for $1\ll r, L$ with $r$ not too close to $L$. For definiteness, we consider the scaling of $S_{\cal I}(r=L/2)$ with $\ln L$~\cite{Gutperle2015, Brehm2015}. Fitting to $S_{\cal I}(r=L/2) = (c/3)\ln L + \tilde{S}_0$~\footnote{Note that for a periodic~(open) system, the coefficient of $\ln L$ is $c/3$~$(c/6)$.} yields the expected central charge $c\simeq0.5$ for all the curves. Recall that for $0\leq b_\sigma<1$, the coefficient of $\ln L$ is $c_{\rm eff}/3$, where the `effective central charge'~$c_{\rm eff}\in [0,1/2)$~\cite{Eisler2010,Brehm2015}. The difference of the offsets, $\tilde{S}_0$, between $b_\epsilon=-1$ and $b_\epsilon=1$ cases is $2\Delta_I=\Delta S(1/2)$ $=-1/2 + \ln2$ or 0 depending on the total system being in a mixed or pure state. For the topological case, both pure and mixed states, the corresponding offset difference is $\Delta_I=\Delta S(1/2)/2=-1/4 + (\ln2)/2$. Importantly, this offset occurs entirely due to a `finite-size effect' correction arising due to the existence of nonlocal zero-modes and bears no relationship to the specific modular S-matrix elements predicted in Refs.~\cite{Gutperle2015, Brehm2015}. 

To summarize, we computed the symmetric and interface EEs for the Ising CFT with a topological defect taking into account the subtle effects of the zero-modes on the EEs. We showed that while both the EEs exhibit identical leading-order logarithmic scaling, the subleading~$O(1)$ corrections are of completely different origin. The subleading term~[$=(\ln 2)/2$]  for the symmetric EE is related to the $g$-function of the corresponding defect at the boundary in the folded picture. However, the corresponding term~[$\Delta S(r/L)/2$] in the interface EE arises only when the subsystem occupies a finite fraction of the total system and is entirely due to the local and nonlocal zero-modes of the topological defect Hamiltonian. In the limit $r\ll L$, {\it there is no additional offset} compared with the case without defect. The interface EE result is in sharp contrast with the existing predictions in terms of the modular S-matrix of the CFT~\cite{Brehm2015, Gutperle2015}, which predict an offset equal to $-\ln 2$ instead. We also computed the EEs in the case of open/free boundary conditions with a defect at the center of the chain~\footnote{See Supplementary Material for details}. The results for symmetric EE were identical to that obtained for the periodic chain. The offset interface EE is the same as for the periodic chain when the total system was pure. The total system being mixed contributes another $(\ln2)/2$ to the offset. 

Several features of the topological defect persist away from the topological point, {\it i.e.,} $b_\sigma\neq1$, and even away from the conformal critical point. We plan to address some of these questions elsewhere. Defect Hamiltonians for other rational CFTs contain more complicated set of zero modes~\cite{Belletete2020}. The question of subleading corrections in EEs for these models remains open. Finally, advancements in measurement of Renyi entropies in engineered quantum systems~\cite{Islam2015, Brydges2019} and quantum simulation~\cite{Monroe2021} can lead to potential verification of our analytical predictions. 

We thank Natan Andrei, Pasquale Calabrese, Christopher Herzog and Ingo Peschel for discussions. We also thank David Rogerson and Frank Pollmann for discussions and collaborations on a related project. AR acknowledges support from a grant from the Simons Foundation (825876, TDN). HS was supported by ERC Advanced Grant NuQFT. 
\bibliography{library_1}
\end{document}